\documentclass[prl,twocolumn,pdftex]{revtex4}
\usepackage{graphicx}
\usepackage{amsmath}

\begin{document}
\title{Effects of Charge Conservation and Flow on Fluctuations of parity-odd Observables at RHIC}
\author{S\"{o}ren Schlichting and Scott Pratt}
\affiliation{Department of Physics and Astronomy and National Superconducting Cyclotron Laboratory,
Michigan State University\\
East Lansing, Michigan 48824}
\date{\today}

\begin{abstract}
The observation of fluctuations of parity-odd angular observables at RHIC has been interpreted as a signal of a local parity violation. We show how the observed correlations can be explained by local charge conservation at freeze-out combined with elliptic flow. Calculations from a blast wave model, which overlays thermal emission onto a collective flow profile,  are shown to account for the experimentally observed signal.
\end{abstract}


\maketitle

The prospect of observing parity violation from the strong interaction in relativistic heavy ion collisions \cite{Kharzeev} has recently gained great attention. Measurements of the STAR collaboration have revealed large fluctuations of parity-odd observables and the signal was proposed to originate from a local parity violation in QCD leading to opposite directions of preferred emission for oppositely charged particles \cite{STAR_parity}. Since the direction of preferred emission fluctuates randomly, the studies are confined to charge-dependent azimuthal angle correlations. The observable measured by STAR is
\begin{equation}
\gamma_{\alpha,\beta}=\frac{\sum_{i \in \alpha,j \in \beta} \cos(\phi_i+\phi_j)}{M_\alpha M_\beta},
\label{gamma_def}
\end{equation}
where $\alpha$ and $\beta$ represent positive or negative charge, $M_\alpha$ and $M_\beta$ are the corresponding multiplicities, the azimuthal angles $\phi$ are measured about the beam axis, and the reaction plane is at $\phi=0$ without loss of generality. In the case where $\alpha$ and $\beta$ refer to the same charge, the $i=j$ terms are excluded from the sum. The proposed source of the parity-odd fluctuations \cite{Kharzeev} is an electric field generated perpendicular to the reaction plane of the initial collision. Whereas the passing ions would be expected to create a non-zero coherent magnetic field for events with non-zero impact parameter, there should per se be no coherent electric field. However, through the anomalous coupling to the parity-odd topological charge in the QCD sector, ${\bf E}_a\cdot{\bf B}_a$, a coherent electric field can be generated, whose direction would fluctuate from being parallel to being anti-parallel with the coherent magnetic field. For each individual nucleon-nucleon collision the sign of ${\bf E}_a\cdot{\bf B}_a$ within each individual flux tube would vary randomly, thus making the generated electric field coherent only in as much as there are several times more particles than flux tubes. The phenomenon is referred to as the ``chiral magnetic effect''. Since one is looking for effects from a coherent electric field that is out-of-plane, the angular correlation should be constructed so that it compares correlations between same-sign and opposite-sign pairs, and so that it compares out-of-plane vs. in-plane correlations. The observable of interest is then
\begin{eqnarray}
\label{eq:gammapdef}
\gamma_P&\equiv&\frac{1}{2}\left(2\gamma_{+-}-\gamma_{++}-\gamma_{--}\right),\\
\nonumber
&=&\frac{4}{M^2}\left\{\sum_{\rm opp.~sign}(\cos\phi_i\cos\phi_j-\sin\phi_i\sin\phi_j)-{\rm ss}\right\},
\end{eqnarray}
where it has been assumed that there are equal numbers of positive and negative charges. This is approximately true for the energies at RHIC. Then $M=M_++M_-$ is the total charged multiplicity, and ``ss'' refers to the corresponding sum with same-sign pairs. A coherent electric field out-of-plane ($\phi=\pi/2$) would give a positive correlation between the $\sin\phi$ terms for same-sign pairs and a negative correlations between the $\sin\phi$ terms for opposite-sign pairs. Thus, $\gamma_P$ would be positive, and is indeed observed to be positive. By subtracting the in-plane correlations, i.e., the $\cos\phi_i\cos\phi_j$ terms, the intent is to eliminate many other sources of angular correlations that are not dependent on the average direction of the pair. Although the chiral magnetic effect would explain the sign of the data, the magnitude of the observed signal is several orders of magnitude higher than some estimates \cite{Asakawa:2010bu,Scott_parity}.

However it has been pointed out that local charge conservation at breakup combined with elliptic flow might explain a major part of the observed signal \cite{Scott_parity}. Similarly cluster particle correlations have been proposed as another alternative explanation \cite{FWang:2010} Charge conservation gives positive correlations between opposite-sign particles resulting in positive terms for $\cos\phi_i\cos\phi_j$  as well as $\sin\phi_i\sin\phi_j$ (opposite to the parity signal). But since there are more such pairs in-plane than out-of-plane due to elliptic flow, and additionally because their correlations are stronger in-plane than out-of-plane (also opposite to the parity signal)  the expression in (\ref{eq:gammapdef}) is dominated by the $\cos\phi$ terms for opposite-sign pairs. Thus this effect also results in positive values of $\gamma_P$. In this letter we investigate how such correlations can reproduce the observed signal by considering a simple blast wave model in which particles are generated according to thermal sources moving with collective flow. By constraining each thermal source to emit equal numbers of positive and negative particles, one can generate non-zero values of $\gamma_P$. The goal of this letter is to quantitatively explore the degree to which the measurement of $\gamma_P$ can be reproduced with a simple thermal model incorporating collective flow and local charge conservation. 

In all fundamental processes charge is created in balancing pairs that are produced at the same point in space time. When the motion is highly collective the correlation in space time translates to a tight correlation of balancing charges in momentum space. This correlation is strongest when charge production takes place late in the collision or diffusion is small \cite{Scott_diff}.  A differential observable that has been exploited for measuring such correlations is the charge balance function \cite{Scott_diff}, which measures the chance that a charge at angle $\phi$ has a balancing charge emitted with angle $\phi+\Delta\phi$. Results from the STAR collaboration have shown that balancing charges are likely to be emitted in a narrow range in rapidity \cite{STAR_old_bf}, consistent with balancing charges being emitted from small neighborhoods such that the relative rapidities are determined mainly by the thermal motion at breakup \cite{balance_models}. Similar results have recently been generated for azimuthal angles \cite{STAR_bf}. Additional contributions to the charge balance funtion due to small-angle correlations from final-state interactions have been considered in \cite{balance_models}, but only affect the results for $\gamma_P$ at the level of a few percent and will therefore be negelected. For equal numbers of positives and negatives, the charge balance function is defined by
\begin{eqnarray}
B(\phi,\Delta\phi)=&&\left(\frac{N_{+-}(\phi,\Delta\phi)-N_{++}(\phi,\Delta\phi)}{dM/d\phi}\right. \nonumber \\
&&+\left.\frac{N_{-+}(\phi,\Delta\phi)-N_{--}(\phi,\Delta\phi)}{dM/d\phi}\right)
\label{eq:bfdef}
\end{eqnarray}
where $N_{\alpha,\beta}(\phi,\Delta\phi)$ is the number of pairs where the type $\alpha$ was emitted at $\phi$ and the type $\beta$ was emitted at $\phi+\Delta\phi$. The like-sign subtraction statistically isolates the balancing partner. The same analysis could be done in the context of correlations, which for this case are identical to balance functions aside from an extra factor of $dM/d\phi$.

The correlation $\gamma_P$ can be expressed in terms of moments of the balance function, after being combined with the angular distribution $dM/d\phi$,
\begin{eqnarray}
\gamma_P&=&\frac{2}{M^2}\int d\phi~d\Delta\phi~\frac{dM}{d\phi} \; B(\phi,\Delta\phi)\nonumber \\
&&\qquad \quad \left[\cos(2\phi)\cos(\Delta\phi) -\sin(2\phi)\sin(\Delta\phi)\right],
\end{eqnarray}
where one has inserted the definition of the balance function (\ref{eq:bfdef}) into the definition for the parity observable (\ref{eq:gammapdef}), and used the angle addition formula, $\cos(\phi_i+\phi_j)=\cos(2\phi_i)\cos(\Delta\phi)-\sin(2\phi_i)\sin(\Delta\phi)$. 

To understand the degree to which charge conservation affects $\gamma_P$, a thermal blast wave model was modified to incorporate local charge conservation. Blast-wave models are simple parameterizations of the breakup configuration. For this study, the model used by STAR to fit elliptic flow data and spectra was employed \cite{STAR_aa}. The model parameters are the breakup temperature $T_{\rm kin}$, the maximum collective velocities in the in-plane and out-of-plane directions, and the spatial anisotropy of the elliptic shape. The elliptic anisotropies were chosen to fit the elliptic flow observable $v_2$, which quantifies the degree to which more particles are emitted in-plane than out-of-plane,
\begin{equation}
v_2\equiv \frac{1}{M}\int d\phi~\frac{dM}{d\phi}\cos(2\phi).
\end{equation}
The anisotropy is driven by the elliptic shape of the initial fireball, as viewed transverse to the beam. The quantity $v_2$ is a staple of RHIC science and has been analyzed both as a function of the centrality of the collision (for very central collisions the initial shape has little anisotropy), as a function of transverse momentum (the anisotropy is stronger for higher $p_t$ particles) and as a function of species type (more massive particles are more sensitive to collective flow) \cite{STAR_aa}. By fitting both spectra and $v_2$, blast wave parameters were determined for several centralities by STAR in \cite{STAR_aa}. In this parametrization the breakup configuration is characterized by four parameters: the kinetic freeze out temperature, the transverse rapidties in-plane and out-of-plane and the spatial anisotropy. The usual method to apply a blast wave model would be to choose a collective velocity consistent with the blast wave parameterization, generate a particle according to the thermal distribution characterized by the temperature and collective velocity, then repeat for several particles. For this calculation, an array of particles is generated rather than a single particle. The array is chosen consistent with a canonical ensemble with a fixed volume (64 fm$^3$) and a chemical temperature, $T_{\rm chem}=175$ MeV \cite{PBM}, so that electric charge, strangeness and baryon number all sum to zero. The particles are then individually assigned momenta according to the (kinetic) breakup temperature and collective flow. In this way, charge conservation is enforced in the most stringent way, with balancing charges being emitted from the same source velocity.
If the charges had been created early and diffused before most of the collective flow developed, balancing particles might have been emitted from regions with significantly different collective velocities. Analysis of charge balance functions as a function of (mid-)rapidity suggest that the emission of balancing charges is indeed highly localized for central events \cite{balance_models}, but less so for peripheral events. Given that the collective velocity gradients in the transverse direction are smaller, the effects of diffusion are expected to be smaller for balance functions binned as a function of azimuthal angle, which makes constraining the balancing charges to originate from sources with the same collective velocity reasonable.

\begin{figure}[ht]
\centerline{\includegraphics[width=0.4\textwidth]{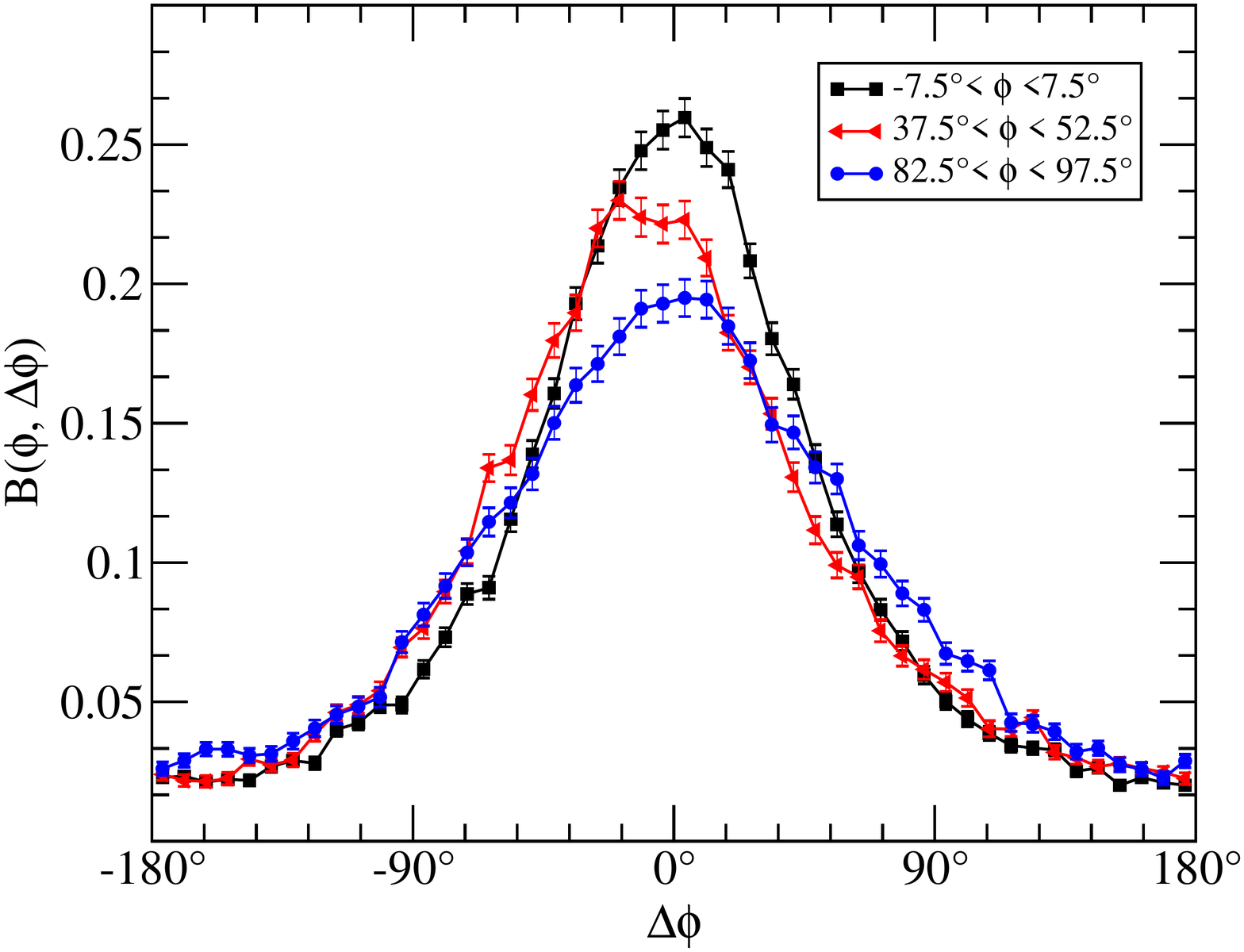}}
\centerline{\includegraphics[width=0.4\textwidth]{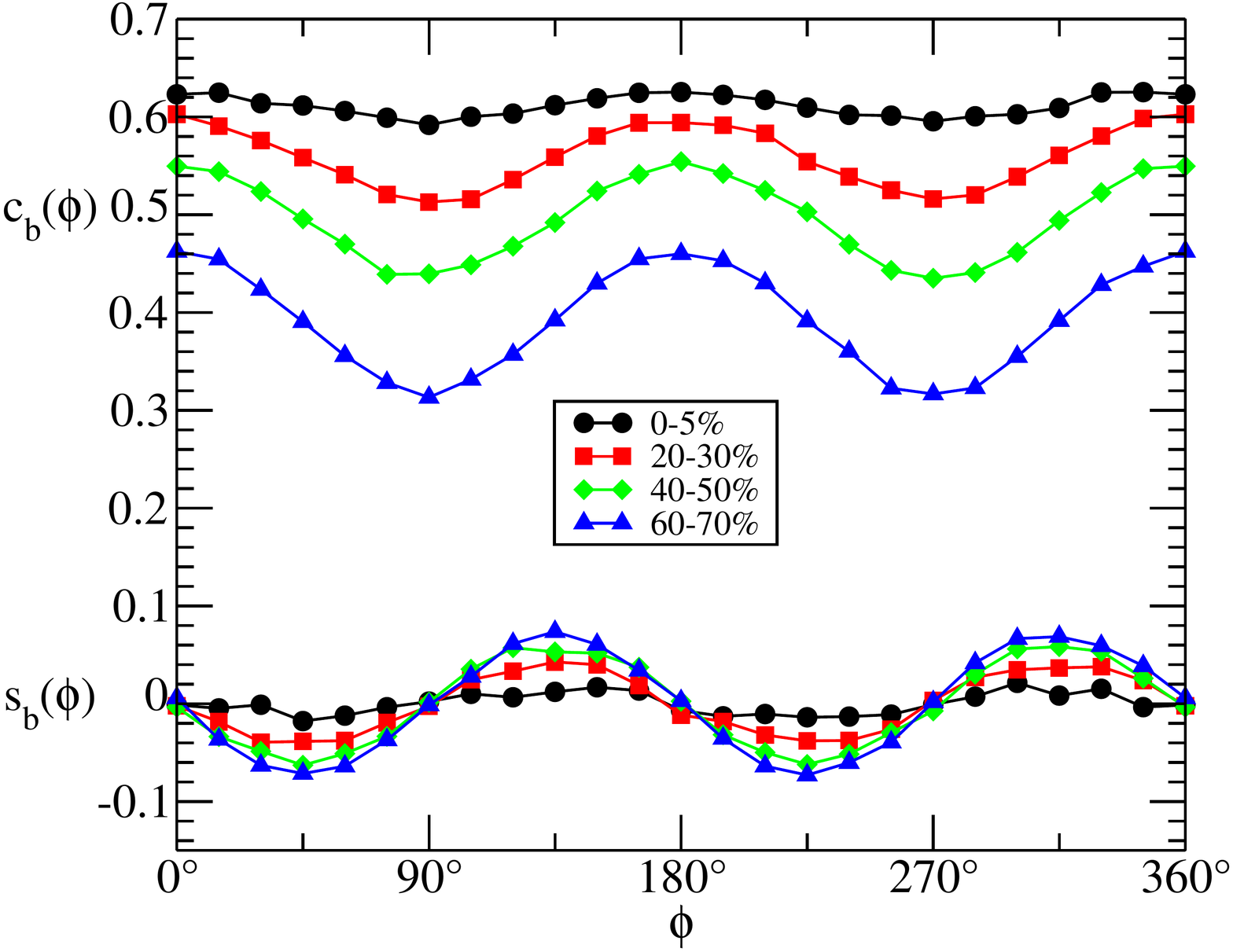}}
\caption{\label{fig:balance}(color online)
UPPER PANEL: Balance function for 40-50\% centrality shown as function of the relative angle included by balancing partners for $\phi=0^\circ$ (black squares), $45^\circ$ (red triangles) and $90^\circ$ (blue circles). LOWER PANEL: The moments of the balance function, $c_b(\phi)$ and $s_b(\phi)$, represent averages of $\cos(\Delta\phi)$ and $\sin(\Delta\phi)$ across the balance function. These are plotted as a function of $\phi$ for various centralities. The structure of $c_b(\phi)$, which is maximized at $\phi=0^\circ,180^\circ$, illustrates how the balance function is narrower for in-plane emission and more central collisions, while the structure of $s_b(\phi)$, which is positive for $\phi=135^\circ,315^\circ$ and negative for $\phi=45^\circ,225^\circ$, shows how balancing charges prefer to be emitted in the in-plane direction. The oscillations increase for more peripheral collisions.}
\end{figure}

Balance functions $B(\phi,\Delta\phi)$ from the blast wave model described above are presented as a function of $\Delta\phi$ in Fig. \ref{fig:balance} for events with centralities of 40-50\% using STAR's parameters \cite{STAR_aa}. Here 0\% centrality corresponds to zero impact parameter and we refer to \cite{STAR_aa} for more details on the classification. The balance function for $\phi=0^\circ$ (in-plane) is narrower than the balance function for $\phi=90^\circ$ (out-of-plane). The stronger focussing of balancing charges derives from the greater collective flow in-plane vs. out-of-plane. For $\phi=45^\circ$, the distribution is biased towards negative values of $\Delta\phi$. This is expected given the elliptic asymmetry, $v_2>0$, which leads to more balancing particles toward the $\phi=0^\circ$ direction as opposed to $\phi=90^\circ$. Depending on which quadrant $\phi$ is located, the balancing charge tends to be found more towards $\phi=0^\circ$ or $\phi=180^\circ$. The lower panel shows the moments of $B(\phi,\Delta\phi)$, 
\begin{eqnarray}
\label{eq:bfmoments}
c_b(\phi)&\equiv&\frac{1}{z_b(\phi)}\int d\Delta\phi~B(\phi,\Delta\phi)\cos(\Delta\phi),\\
\nonumber
s_b(\phi)&\equiv&\frac{1}{z_b(\phi)}\int d\Delta\phi~B(\phi,\Delta\phi)\sin(\Delta\phi).
\end{eqnarray}
where 
\begin{eqnarray}
z_b(\phi)&\equiv&\int d\Delta\phi~B(\phi,\Delta\phi),
\end{eqnarray}
is the normalization of the balance function and represents the probability of detecting the balancing charge given the observation of a charge at $\phi$. It would be unity for a perfect detector, but is reduced by both the finite acceptance and efficiency of the experiment. The quantity $c_b(\phi)$ determines the width of the balance function and would be unity for a very narrow balance function whereas it vanishs in the case where the balancing charges were emitted randomly. The quantity $s_b(\phi)$ measures the degree to which the balance function is asymmetric under a reflection symmetry of $\Delta\phi \rightarrow -\Delta\phi$. For pairs around $\phi=45^\circ$ this corresponds to the probability for the balancing charge to be emitted in in-plane direction vs. in out-of-plane direction.

\begin{figure}[ht]
\centerline{
	\includegraphics[width=0.4\textwidth]{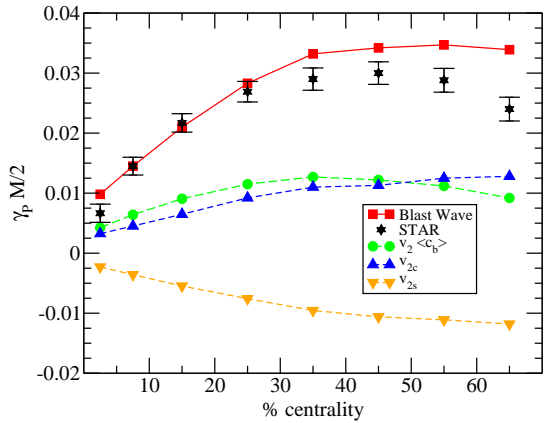}
}
\caption{\label{fig:gamma}(color online)
	Parity observable from STAR (black stars) and blast wave calculations (red squares). The three contributions to the signal are defined in Eq. (\ref{eq:components}) and are plotted with dashed lines. $v_2\langle c_b\rangle$ (green circles) derives from having more balancing pairs in-plane than out-of-plane while $v_{2,c}$ (blue triangles) quantifies the degree to which in-plane pairs are more tightly correlated than out-of-plane pairs. $v_{2,s}$ (orange triangles) reflects that the balancing charge is more likely to be found towards the event plane. 
}
\end{figure}

The parity observable $\gamma_P$ can be expressed, using the moments of the balance function defined in (\ref{eq:bfmoments}),
\begin{eqnarray}
\label{eq:components}
\gamma_P&=&\frac{2}{M}\left[v_2\langle c_b(\phi)\rangle+v_{2c}-v_{2s}\right],
\end{eqnarray}
where we introduced
\begin{eqnarray}
v_{2c}&\equiv&\langle c_b(\phi)\cos(2\phi)\rangle-v_2\langle c_b(\phi)\rangle, \nonumber \\
v_{2s}&\equiv&\langle s_b(\phi)\sin(2\phi)\rangle,\nonumber \\
\langle f(\phi)\rangle &\equiv& \frac{1}{M}\int d\phi~\frac{dM}{d\phi}~z_b(\phi) f(\phi).
\end{eqnarray}
The three contributions to $\gamma_P$ derive from: a) having more balancing pairs in-plane than out-of-plane $(v_2\langle c_b\rangle)$, b) having the in-plane pairs being more tightly correlated in $\Delta\phi$ than the out-of-plane pairs $(v_{2c})$and c) having the balancing charge more likely being emitted towards the event plane $(v_{2s})$. The first term was estimated in \cite{Scott_parity} to be a significant fraction of the observed signal, but the latter two terms could not be estimated without a more detailed model like this one. The contributions for $\gamma_P$ obtained from the blast wave calculation are displayed in Fig. \ref{fig:gamma}. For better visibility the values are scaled by the multiplicity (for the STAR data we use the experimental multiplicity \cite{GDWmult} to account for efficiency and acceptance). To compare to the STAR data \cite{STAR_parity} we use the same acceptance cuts in transverse momentum and pseudorapidity and we assume perfect detector efficiency for the blast wave calculation. The necessary efficiency correction is done by rescaling the results to reproduce the experimental normalization of the balance function \cite{STAR_bf}, i.e. we multiply the expressions for $\langle c_b(\phi) \rangle$, $v_{2c}$ and $v_{2s}$ by the ratio of experimental to blast wave normalization.

For central collisions the STAR data is well reproduced by the blast wave model, however for more peripheral collisions the model produces higher correlations than experimentally observed. As we assumed that balancing particles are emitted with the same collective flow, the correlations from the blast wave model shown in Fig. \ref{fig:gamma}, are the strongest possible contribution to the signal induced by charge conservation. Consequently the model is expected to over-predict the measured signal. Whereas for central collisions this assumption has been shown to be consistent with measured balance functions \cite{STAR_bf}, the locality of charge conservation at breakup appears to be less exact for peripheral collisions. The STAR analysis of charge balance functions \cite{STAR_bf}, shows a significant broadening of charge balance functions for larger impact parameters, a part of which is expected from the higher kinetic freeze-out temperature and less transverse collective flow. By relaxing the stringent conditions of having the balancing charges emitted from the same azimuthal angle, the balance functions will become wider and the accompanying correlation $\gamma_P$ should move down toward the data. In the limit that the balancing charge were allowed to come from any point in the blast wave, the contribution to $\gamma_P$ would vanish.

The calculations presented here demonstrate that local charge conservation overlaid with elliptic flow can readily explain the difference between the opposite-sign and same-sign correlations seen by STAR. The over-simplified picture of the decoupling provided by the  blast wave model is rather crude and in some ways the parameterization is arbitrary, but should be accurate to within a few tens of percent. To better model how the observed $\gamma_P$ is induced by these mechanisms, one should employ a more detailed model of both the collision dynamics and decoupling and of the correlation in space-time between balancing charges. We emphasize that the mechanism described here only explains the difference between the opposite-sign and same-sign correlations. This is sufficient to show that large parity fluctuations are not warranted by the data, although the strong same sign-correlations observed at STAR \cite{STAR_parity} have yet to be explained. It was suggested in \cite{Scott_parity} that these might be induced by momentum conservation, however there is so far no quantitative theoretical estimate of this effect. A more completely satisfying model would describe both the same-sign and opposite-sign correlations independently, and would also reproduce the  more detailed differential correlations, such as charge balance functions, in addition to the integrated correlations, $\gamma_{\alpha\beta}$. We expect such analyses to be pursued in the near future.

This work was supported by the U.S. Department of Energy, Grant No. DE-FG02-03ER41259. The authors thank Gary Westfall, Hui Wang and Terence Tarnowsky for insightful discussions.


\begin{thebibliography}{99}
\bibitem{GDWmult} Experimental multiplicities were generously provided by the STAR collaboration.
\end{thebibliography}
\end{document}